\documentclass[prb,twocolumn,showpacs]{revtex4}
\usepackage{graphicx} 
\usepackage{amsmath,amsthm,amssymb,mathrsfs}
\usepackage{bm}

\begin{document}

\title{Instability analysis of spin-torque oscillator with an in-plane magnetized free layer and a perpendicularly magnetized pinned layer}

\author{Tomohiro Taniguchi and Hitoshi Kubota}
 \affiliation{
 National Institute of Advanced Industrial Science and Technology (AIST), Spintronics Research Center, Tsukuba, Ibaraki 305-8568, Japan 
 }

 \begin{abstract}
{
We study the theoretical conditions to excite a stable self-oscillation in a spin-torque oscillator 
with an in-plane magnetized free layer and a perpendicularly magnetized pinned layer 
in the presence of magnetic field pointing in an arbitrary direction. 
The linearized Landau-Lifshitz-Gilbert (LLG) equation is found to be inapplicable to evaluate the threshold between the stable and self-oscillation states 
because the critical current density estimated from the linearized equation is considerably larger than that found in the numerical simulation. 
We derive a theoretical formula of the threshold current density by focusing on the energy gain of the magnetization from the spin torque 
during a time shorter than a precession period. 
A good agreement between the derived formula and the numerical simulation is obtained. 
The condition to stabilize the out-of-plane self-oscillation above the threshold is also discussed. 
}
 \end{abstract}

 \pacs{75.78.Jp, 75.76.+j, 85.75.-d}
 \maketitle




\section{Introduction}
\label{sec:Introduction}

A spin polarized current injected into a nanostructured ferromagnet creates spin torque 
through the spin-transfer effect [\onlinecite{slonczewski96,berger96,slonczewski05}]. 
The spin torque provides a rich variety of magnetization dynamics 
such as switching or self-oscillation [\onlinecite{katine00,kiselev03,rippard04,kubota05,krivorotov05,kubota13,tamaru14}]. 
In particular, a spin-torque oscillator consisting of an in-plane magnetized free layer and a perpendicularly magnetized pinned layer
has been an attractive research subject in the field of magnetism. 
[\onlinecite{kent04,lee05,zhu06,houssameddine07,firastrau07,ebels08,silva10,suto12,lacoste13,kudo14,bosu16}]. 
In this type of spin-torque oscillator, the spin torque forces the magnetization of the free layer into out of plane, 
and excites a large amplitude oscillation around the perpendicular axis. 
A high symmetry along the perpendicular direction in this system makes it easy 
to investigate the oscillation properties theoretically [\onlinecite{silva10}]. 
In order to observe the oscillation experimentally through magnetoresistance effect, however, the symmetry breaking should occur 
since the change of the relative angle between the magnetizations of the free and pinned layers in time is necessary. 
The linear analysis in the presence of an in-plane anisotropy [\onlinecite{ebels08}] 
or the perturbation approach to the system additionally having an in-plane magnetized reference layer [\onlinecite{lacoste13}] have been made to develop practical theory. 


The application of an external magnetic field tilted from the perpendicular axis also breaks the symmetry 
and enables us to measure the oscillation experimentally. 
In other geometries, the experimental studies have shown that the oscillation properties such as 
the threshold current to excite the self-oscillation strongly depend on the field direction [\onlinecite{rippard04,tamaru14}]. 
On the other hand, the role of the magnetic field on the self-oscillation properties in this geometry has not been fully understood yet. 
For example, it is still unclear how much current is necessary to excite the out-of-plane self-oscillation 
in the presence of the magnetic field pointing in an arbitrary direction, 
while it is known that infinitesimal current can excite the self-oscillation for the highly symmetric case [\onlinecite{lee05,silva10}]. 




In this paper, we investigate theoretical conditions to excite the self-oscillation 
in a spin-torque oscillator with an in-plane magnetized free layer and a perpendicularly magnetized pinned layer 
in the presence of an external magnetic field. 
We solve the Landau-Lifshitz-Gilbert (LLG) equation both numerically and analytically. 
The main findings in this paper are as follows. 
First, we find that the linearized LLG equation is no longer useful to evaluate the instability threshold in the present system. 
The critical current density evaluated from the linearized LLG equation is two orders of magnitude larger than 
the instability threshold estimated from the numerical simulation. 
Second, we derive the theoretical formula determining the instability threshold. 
The main difference between the linear analysis and our result is that 
when a periodic precession around the stable state is assumed in the linear analysis, 
while we focus on the transition of the magnetization from the stable state to the out-of-plane self-oscillation state 
during a time shorter than the precession period. 
A good agreement between the numerical simulation and our formula is obtained. 
Third, we derive theoretical conditions to guarantee the present results, 
i.e., the condition that our formula of the threshold current density 
works better than the linear analysis to estimate the instability threshold, 
and the condition to stabilize the out-of-plane self-oscillation. 


This paper is organized as follows. 
In Sec. \ref{sec:Numerical simulation and linear analysis}, 
we show the numerical simulation results near the instability of the initial state. 
We also solve the linearized LLG equation analytically. 
In Sec. \ref{sec:Theoretical formula of threshold current}, 
we derive a theoretical formula of the threshold current, 
and confirm its validity by comparing the results obtained from the formula with the numerical simulation. 
The conclusion is summarized in Sec. \ref{sec:Conclusion}. 



\begin{figure*}
\centerline{\includegraphics[width=2.0\columnwidth]{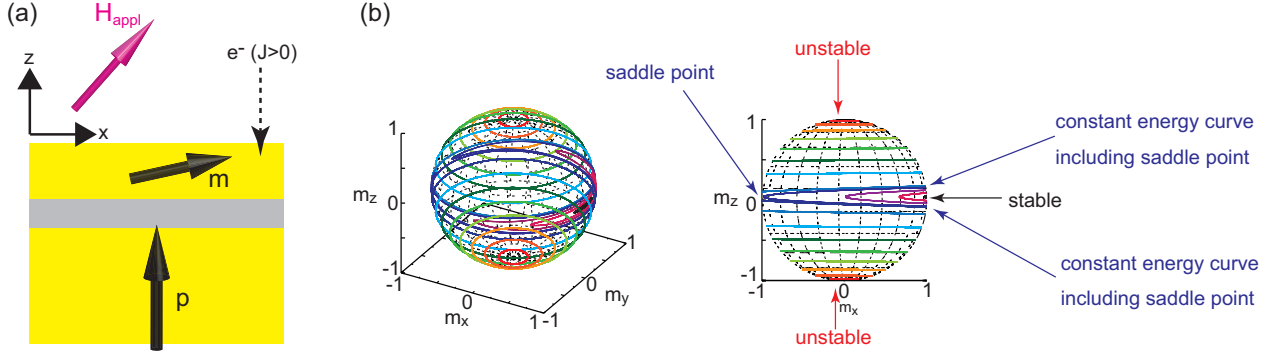}}
\caption{
         (a) Schematic view of the system considered in this study. 
             The unit vectors pointing in the magnetization direction of the free and pinned layers are denoted as $\mathbf{m}$ and $\mathbf{p}$, respectively. 
             The positive electric current corresponds to the electrons flowing from the free layer to the pinned layer. 
             The external field lies in the $xz$ plane. 
         (b) Schematic views of the constant energy curves. 
         \vspace{-3ex}}
\label{fig:fig1}
\end{figure*}



\section{Numerical simulation and linear analysis}
\label{sec:Numerical simulation and linear analysis}

In this section, we investigate the threshold current density 
which is necessary to destabilize the magnetization in the stable state 
by numerically solving the LLG equation. 
We also compare the numerical result with the analytical values of the critical current density $j_{\rm c}$ estimated from the linearized LLG equation. 
Throughout this paper, the term "threshold current" indicates the current destabilizing the stable state 
calculated in the numerical simulation or from the formula which is also well consistent with the numerical simulation, 
while the term "critical current" is a current estimated from the linearized LLG equation. 


\subsection{System description}

The system we consider is schematically shown in Fig. \ref{fig:fig1}(a). 
The $z$ axis is perpendicular to the film plane. 
The unit vectors pointing in the magnetization direction of the free and pinned layers are denoted as $\mathbf{m}$ and $\mathbf{p}$, respectively. 
The magnetization of the pinned layer points to the positive $z$ direction, $\mathbf{p}=+\mathbf{e}_{z}$. 
The positive current is defined as the electrons flowing from the free layer to the pinned layer. 
We use the macrospin approximation to the free layer. 
The magnetization dynamics is described by the LLG equation 
\begin{equation}
  \frac{d \mathbf{m}}{d t}
  =
  -\gamma
  \mathbf{m}
  \times
  \mathbf{H}
  -
  \gamma
  H_{\rm s}
  \mathbf{m}
  \times
  \left(
    \mathbf{p}
    \times
    \mathbf{m}
  \right)
  +
  \alpha
  \mathbf{m}
  \times
  \frac{d \mathbf{m}}{d t},
  \label{eq:LLG}
\end{equation}
where $\gamma$ and $\alpha$ are the gyromagnetic ratio and the Gilbert damping constant, respectively. 
We use the approximation $1+\alpha^{2} \simeq 1$ 
because the damping constant for typical ferromagnets is on the order of $10^{-2}-10^{-3}$ [\onlinecite{oogane06,tsunegi14}]. 
The spin-torque strength is 
\begin{equation}
  H_{\rm s}
  =
  \frac{\hbar \eta j}{2eMd}, 
  \label{eq:H_s}
\end{equation}
where $\eta$ is the spin polarization of the electric current density $j$, 
while $M$ and $d$ are the saturation magnetization and the thickness of the free layer, respectively. 
We neglect the asymmetry of the spin torque described by the term $1/(1+\lambda \mathbf{m}\cdot\mathbf{p})$ [\onlinecite{slonczewski96}] here, for simplicity. 
The critical current density in the presence of this factor, as well as its role, is briefly summarized in Appendix \ref{sec:AppendixA}. 
The magnetic field $\mathbf{H}$ consists of the demagnetization field along the $z$ direction, $-4\pi M$, 
and the applied $\mathbf{H}_{\rm appl}$ expressed as 
\begin{equation}
  \mathbf{H}
  =
  \mathbf{H}_{\rm appl}
  -
  4\pi M 
  m_{z} 
  \mathbf{e}_{z}.
  \label{eq:field}
\end{equation}
The applied field $\mathbf{H}_{\rm appl}$ is tilted from the $z$ axis and assumed to lie in the $xz$ plane for convention, 
i.e., 
\begin{equation}
  \mathbf{H}_{\rm appl}
  =
  H_{\rm appl}
  \sin\theta_{H}
  \mathbf{e}_{x}
  +
  H_{\rm appl}
  \cos\theta_{H}
  \mathbf{e}_{z},
  \label{eq:applied_field}
\end{equation}
where $H_{\rm appl}$ and $\theta_{H}$ are the amplitude and the tilted angle from the $z$ axis of the applied field, respectively. 
The magnetic field relates to the energy density $E$ via $E=-M \int d \mathbf{m}\cdot\mathbf{H}$, 
which in the present system is 
\begin{equation}
\begin{split}
  E
  =&
  -M H_{\rm appl}
  \left(
    \sin\theta_{H}
    m_{x}
    +
    \cos\theta_{H}
    m_{z}
  \right)
  +
  2\pi M^{2}
  m_{z}^{2}.
  \label{eq:energy}
\end{split}
\end{equation}
Here, we assume that $H_{\rm appl}<4\pi M$, 
and therefore, the stable state, i.e., the minimum of Eq. (\ref{eq:energy}), locates close to the $x$ axis. 
Note that the magnetization dynamics described by Eq. (\ref{eq:LLG}) can be regarded as the motion of a point particle on a unit sphere. 


The values of the parameters used in this section are brought from typical experiments [\onlinecite{hiramatsu16}], 
$M=1300$ emu/c.c., 
$\gamma=1.764 \times 10^{7}$ rad/(Oe s), 
$\alpha=0.01$, 
$d=2$ nm, 
and $\eta=0.5$. 
The magnitude of the applied field is $H_{\rm appl}=650$ Oe, 
while the field angle is $\theta_{H}=5^{\circ}$. 
Figure \ref{fig:fig1}(b) shows the constant energy curves of Eq. (\ref{eq:energy}) with these parameters. 
Note that the stable (minimum energy) state, the saddle point, and the unstable (local maximum) states of the energy density $E$ all exist in the $xz$ plane. 
The stable state locates in the positive $x$ region, while the saddle point exists in the negative $x$ region. 
Also, the unstable states slightly shift from the $z$ axis due to the applied field. 
We denote the energies corresponding to the stable state, the saddle point, and the unstable states as 
$E_{\rm min}$, $E_{\rm saddle}$, and $E_{\rm max \pm}$, 
where the subscript $\pm$ distinguishes the unstable states in the positive ($+$) and negative ($-$) $z$ region. 
For $\theta_{H} \neq 90^{\circ}$, $E_{\rm max+} \neq E_{\rm max-}$. 
The constant energy curves in Fig. \ref{fig:fig1}(b) are classified to the ellipses around the $x$ and $z$ axes. 
The energy density $E$ corresponding to the curves around the $x$ axis is in the region of $E_{\rm min} \le E \le E_{\rm saddle}$, 
while that for the curves around the $z$ axis is in the region of $E_{\rm saddle} < E \le E_{\rm max \pm}$. 


\subsection{Linear analysis}

The conventional method to estimate the minimum current density to destabilize the stable state 
is linearizing the LLG equation 
and investigating the oscillating solution of the magnetization with a complex frequency [\onlinecite{sun00,grollier03,morise05}]. 
In this section, we derive the theoretical formula of the critical current density and estimate its value. 

We introduce the zenith and azimuth angles $(\theta,\varphi)$ as 
$\mathbf{m}=(\sin\theta\cos\varphi,\sin\theta\sin\varphi,\cos\theta)$ 
to identify the magnetization direction. 
In particular, the angles corresponding to the stable state are denoted as $(\theta_{0},\varphi_{0})$. 
In the present case, $\varphi_{0}=0$, 
and $\theta_{0}$ is determined by the condition $(\partial E/\partial\theta)_{\varphi=\varphi_{0}}=0$, 
\begin{equation}
  H_{\rm appl}
  \sin(\theta_{H}-\theta_{0})
  +
  4\pi M 
  \sin\theta_{0}
  \cos\theta_{0}
  =
  0.
\end{equation}
We introduce a new coordinate $XYZ$ where the $Z$ axis is parallel to the magnetization in the stable state $(\theta_{0},\varphi_{0})$. 
A small amplitude oscillation of the magnetization around a stationary point is described by 
the following linearized LLG equation 
(the detail of the derivation is shown in Appendix \ref{sec:AppendixA}) 
\begin{equation}
\begin{split}
  &
  \frac{1}{\gamma}
  \frac{d}{dt}
  \begin{pmatrix}
    m_{X} \\
    m_{Y}
  \end{pmatrix}
  +
  \mathsf{M}
  \begin{pmatrix}
    m_{X} \\
    m_{Y} 
  \end{pmatrix}
  =
  H_{\rm s}
  \begin{pmatrix}
    \sin\theta_{0} \\
    0 
  \end{pmatrix}, 
  \label{eq:linearized_LLG}
\end{split}
\end{equation}
where 
\begin{equation}
  \mathsf{M}
  =
  \begin{pmatrix}
    \alpha H_{X} - H_{\rm s} \cos\theta_{0} & H_{Y} \\
    -H_{X} & \alpha H_{Y} - H_{\rm s} \cos\theta_{0}
  \end{pmatrix}
\end{equation}
with $H_{X}=H_{\rm appl}\cos(\theta_{H}-\theta_{0})-4\pi M \cos 2\theta_{0}$ and $H_{Y}=H_{\rm appl}\cos(\theta_{H}-\theta_{0})-4\pi M \cos^{2}\theta_{0}$. 
The solution of Eq. (\ref{eq:linearized_LLG}) has a form of 
$\exp\{\gamma[ \pm i \sqrt{{\rm det}[\mathsf{M}]-({\rm Tr}[\mathsf{M}/2])^{2}} - {\rm Tr}[\mathsf{M}]/2]t\}$. 
The critical current density is defined as the current density satisfying 
${\rm Re}[\pm i \sqrt{{\rm det}[\mathsf{M}]-({\rm Tr}[\mathsf{M}/2])^{2}} - {\rm Tr}[\mathsf{M}]/2]=0$. 
For a small $\alpha$, this condition is approximated to ${\rm Tr}[\mathsf{M}/2]=0$ 
because $({\rm Tr}[\mathsf{M}/2])^{2}/{\rm det}[\mathsf{M}] \sim \alpha^{2} \simeq 0$. 
Therefore, the critical current density becomes 
\begin{equation}
  j_{\rm c}
  =
  \frac{2 \alpha eMd}{\hbar \eta \cos\theta_{0}}
  \left[
    H_{\rm appl}
    \cos(\theta_{H}-\theta_{0})
    -
    4\pi M 
    \frac{\cos^{2}\theta_{0} + \cos 2 \theta_{0}}{2}
  \right].
  \label{eq:jc}
\end{equation}
Substituting the above parameters, 
we find that $\theta_{0} \simeq 87.7^{\circ}$ and $j_{\rm c}=328 \times 10^{6}$A/cm${}^{2}$. 



\subsection{Numerical simulation}

\begin{figure}
\centerline{\includegraphics[width=1.0\columnwidth]{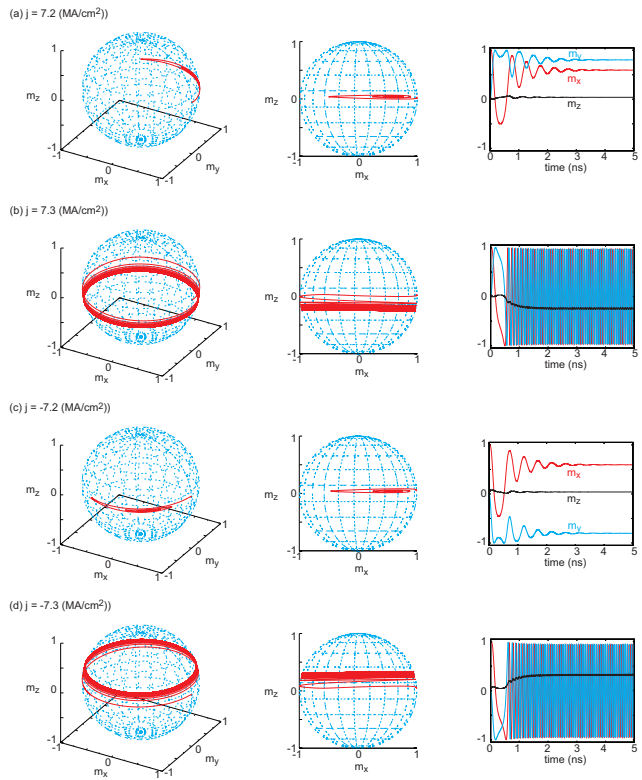}}
\caption{
         The trajectories of the magnetization dynamics on the unit spheres. 
         The time evolutions of the magnetization components are also shown. 
         The values of the current density $j$ are (a) 7.2, (b) 7.3, (c) -7.2, and (d) -7.3 $\times 10^{6}$ A/cm${}^{2}$. 
         \vspace{-3ex}}
\label{fig:fig2}
\end{figure}

Figures \ref{fig:fig2}(a)-(d) show the magnetization dynamics on the unit sphere 
and time developments of the components of $\mathbf{m}$, 
obtained by numerically solving the LLG equation, Eq. (\ref{eq:LLG}). 
The current density is (a) 7.2, (b) 7.3, (c) -7.2, and (d) -7.3 $\times 10^{6}$ A/cm${}^{2}$. 
As shown, when the current magnitude $|j|$ is smaller than $7.2 \times 10^{6}$ A/cm${}^{2}$, 
the magnetization finally moves to another point and stops its dynamics. 
On the other hand, the magnetization shows the self-oscillation for $|j| \ge 7.3 \times 10^{6}$A/cm${}^{2}$. 
The $z$ component of the magnetization moves to the positive (negative) $z$ direction for the negative (positive) current 
because the negative (positive) current prefers $\mathbf{m}$ to be parallel (antiparallel) to the magnetization of the pinned layer, $\mathbf{p}=+\mathbf{e}_{z}$.



Three important conclusions are obtained from Fig. \ref{fig:fig2}. 
First, the threshold current density to destabilize the initial stable state, $\simeq \pm 7.3 \times 10^{6}$ A/cm${}^{2}$, 
is two orders of magnitude smaller than the critical current density, $j_{\rm c}=328 \times 10^{6}$ A/cm${}^{2}$, estimated from the linearized LLG equation.
Second, both positive and negative currents can destabilize the initial state, 
while the sign of $j_{\rm c}$ is fixed (positive for $\theta_{H}<90^{\circ}$). 
Third, the magnetization precesses around the $z$ axis above the threshold. 
Note that the self-oscillation occurs on the trajectory close to the constant energy curve. 
Although the energy landscape has the constant energy curves around the $x$ axis, as shown in Fig. \ref{fig:fig1}(b), 
an in-plane precession around the $x$ axis does not appear. 
In the next section, we explain the physical meanings of such behavior. 



\section{Theoretical formula of threshold current}
\label{sec:Theoretical formula of threshold current}

The results discussed in the previous section indicates that 
the linear analysis is no longer applicable to evaluate the instability threshold, 
although the linear analysis has been widely used to analyze the spin torque induced magnetization dynamics [\onlinecite{sun00,grollier03,morise05}]. 
In this section, we clarify the reason for the breakdown of the linear analysis, 
and derive a theoretical formula of the threshold current density by focusing on 
the energy gain of the free layer generated from the work done by spin torque. 


\subsection{LLG equation averaged over constant energy curves}

Here, let us discuss the averaging technique of the LLG equation on the constant energy curves. 
This method has been used in several works to analyze the self-oscillation and the thermally activated magnetization switching induced by spin torque, 
the microwave assisted magnetization reversal, and so on 
[\onlinecite{bertotti04,bertotti05,serpico05,bertotti09,apalkov05,hillebrands06,bazaliy11,dykman12,newhall13,taniguchi13PRB,taniguchi14,taniguchi15,pinna14}]. 
As will be discussed below, the critical current density $j_{\rm c}$ introduced above corresponds to a special limit of this averaging technique. 
Therefore, by reviewing the derivation of the averaged LLG equation, 
the reason why the linearized LLG equation does not work to estimate the instability condition accurately will be clarified. 

The self-oscillation is a steady precession on a constant energy curve of $E$ excited by the magnetic field torque ($-\gamma\mathbf{m}\times\mathbf{H}$). 
To maintain the precession, the spin torque should balance with the damping torque. 
Note however that the spin torque and the damping torque have different angular dependences. 
Therefore, strictly speaking, 
the spin torque may overcome the damping torque at certain points on the precession trajectory, 
the damping torque may however overcome the spin torque at other points. 
The self-oscillation is maintained when the shift from the constant energy curve due to the imbalance between the spin torque and the damping torque is sufficiently small. 
In that case, the magnetization can return back to the original constant energy curve during the precession. 
When this condition is satisfied, we obtain the following averaged LLG equation, 
\begin{equation}
  \oint 
  dt 
  \frac{dE}{dt}
  =
  \mathscr{W}_{\rm s}(E)
  +
  \mathscr{W}_{\alpha}(E),
  \label{eq:averaged_LLG}
\end{equation}
where the integral range is a precession period on a constant energy curve of $E$. 
The work done by spin torque and the dissipation due to the damping during the precession are 
\begin{equation}
  \mathscr{W}_{\rm s}
  =
  \oint dt 
  \gamma 
  M H_{\rm s}
  \left[
    \mathbf{p}
    \cdot
    \mathbf{H}
    -
    \left(
      \mathbf{m}
      \cdot
      \mathbf{p}
    \right)
    \left(
      \mathbf{m}
      \cdot
      \mathbf{H}
    \right)
  \right],
  \label{eq:Melnikov_s}
\end{equation}
\begin{equation}
  \mathscr{W}_{\alpha}
  =
  -\oint dt 
  \alpha
  \gamma
  M
  \left[
    \mathbf{H}^{2}
    -
    \left(
      \mathbf{m}
      \cdot
      \mathbf{H}
    \right)^{2}
  \right],
  \label{eq:Melnikov_alpha}
\end{equation}
respectively. 
Since the energy density averaged over the precession is conserved in the self-oscillation state, 
the self-oscillation is described by the equation $\oint dt (dE/dt)=0$. 
Therefore, the current density necessary to excite a self-oscillation on a certain constant energy curve of $E$ is 
\begin{equation}
  j(E)
  =
  \frac{2 \alpha eMd}{\hbar \eta}
  \frac{\oint dt [\mathbf{H}^{2}-(\mathbf{m}\cdot\mathbf{H})^{2}]}{\oint dt [\mathbf{p}\cdot\mathbf{H}-(\mathbf{m}\cdot\mathbf{p})(\mathbf{m}\cdot\mathbf{H})]}.
  \label{eq:balance_current}
\end{equation}
The explicit form of $j(E)$ for an arbitrary $E$ is obtained, in principle, 
by substituting the solution of the precession trajectory on a constant energy curve, 
which is described by $d \mathbf{m}/dt=-\gamma\mathbf{m}\times\mathbf{H}$. 
However, the solution is hardly obtained because the equation is a nonlinear equation. 
Therefore, we evaluate the integrals in Eq. (\ref{eq:balance_current}) numerically, 
except for special cases mentioned below (see also Appendix \ref{sec:AppendixB}). 
The technique to evaluate the integrals in Eq. (\ref{eq:balance_current}) is shown, for example, in Ref. [\onlinecite{taniguchi14}]. 
The damping constant $\alpha$ is assumed to be scalar in the above formulation. 
On the other hand, a tensor damping was proposed in Ref. [\onlinecite{safonov02}]. 
The presence of the tensor damping was also suggested in the spin-torque problem [\onlinecite{zhang09}]. 
The effect of the tensor damping can be taken into account by replacing $\alpha$ in Eq. (\ref{eq:Melnikov_alpha}) with the tensor damping; 
see Appendix C of Ref. [\onlinecite{taniguchi14}].




\begin{figure}
\centerline{\includegraphics[width=1.0\columnwidth]{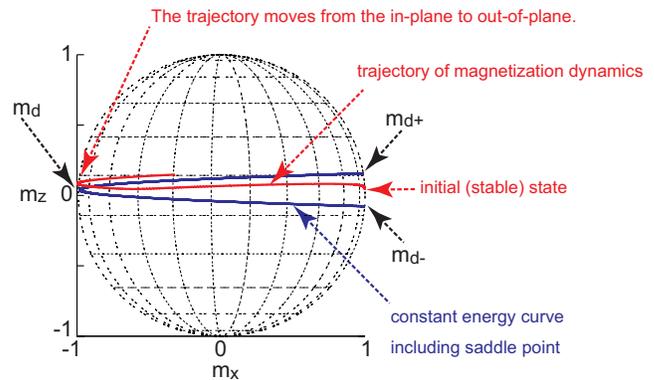}}
\caption{
         The constant energy curve including the saddle point 
         and the trajectory of the magnetization dynamics for $j=-7.3 \times 10^{6}$ A/cm${}^{2}$. 
         The point $\mathbf{m}_{\rm d}$ is the saddle point, 
         while $\mathbf{m}_{\rm d\pm}$ are points on the constant energy curve of $E_{\rm saddle}$ 
         and locating in the $xz$ plane. 
         \vspace{-3ex}}
\label{fig:fig3}
\end{figure}



\subsection{Derivation of threshold current}

Note that the critical current density $j_{\rm c}$, Eq. (\ref{eq:jc}), obtained from the linearized LLG equation relates to Eq. (\ref{eq:balance_current}) via 
\begin{equation}
  j_{\rm c}
  =
  \lim_{E \to E_{\rm min}}
  j(E). 
\end{equation}
Therefore, the fact that the critical current density $j_{\rm c}$ 
is quite larger than the threshold current density found in the numerical simulation 
indicates the breakdown of applying averaged LLG equation. 


An important assumption in the averaged LLG equation is that the magnitudes of the spin torque and the damping torque are sufficiently small. 
Thus, a shift of the magnetization from a constant energy curve due to the imbalance between these torques is also small. 
However, this assumption is not satisfied in the present case. 
Figure \ref{fig:fig3} shows the trajectory of the magnetization dynamics obtained from the numerical simulation,
where the current density $j=-7.3 \times 10^{6}$ A/cm${}^{2}$ is the threshold value found in Fig. \ref{fig:fig2}(d). 
We also show the constant energy curve including the saddle point $\mathbf{m}_{\rm d}$. 
We should remind the readers that there are two kinds of constant energy curves, as shown in Fig. \ref{fig:fig1}(b), 
i.e., the curves around the $x$ axis corresponding to $E_{\rm min} \le E \le E_{\rm saddle}$ 
and the curves around the $z$ axis corresponding to $E_{\rm saddle} < E \le E_{\rm max \pm}$. 
The constant energy curve of $E_{\rm saddle}$ separates these in-plane and out-of-plane regions. 
As shown in Fig. \ref{fig:fig3}, while the magnetization moves from the initial state to a point close to the saddle point, 
the magnetization crosses the constant energy curves of $E_{\rm saddle}$, 
and transfers from the in-plane region to the out-of-plane region. 
A periodic oscillation around the stable state ($x$ axis) is not excited. 
This result is the evidence that the assumption used in Eq. (\ref{eq:balance_current}), as well as Eq. (\ref{eq:jc}), is broken. 
Therefore, the critical current density $j_{\rm c}$ does not work to estimate the instability of the magnetization around the stable state accurately. 


The inapplicability of the linearized LLG equation also relates to the value of the damping constant $\alpha$. 
Note that both the spin torque and the damping torque move the magnetization from a constant energy curve, 
by either supplying or dissipating the energy from the free layer. 
Therefore, the averaging technique of the LLG equation, as well as the linearization of the LLG equation, works well for low damping case. 
The fact that the linearized LLG equation could not be applied in the above numerical simulation indicates that 
the value of the damping constant in the present system is high and that the precession around the stable state is not stabilized. 
The range of the damping constant where the linearized LLG equation will be applicable is discussed in Sec. \ref{sec:Applicability of the present theory} below. 


Figure \ref{fig:fig3} suggests that the magnetization can climb up the energy barrier $E_{\rm saddle}-E_{\rm min}$ 
by absorbing energy due to the work done by the spin torque 
during a time shorter than a precession period around the stable state. 
Therefore, the threshold current density can be defined as a current density satisfying the following equation, 
\begin{equation}
  \int_{\mathbf{m}_{\rm min}}^{\mathbf{m}_{\rm d}} dt 
  \frac{dE}{dt}
  =
  E_{\rm saddle}
  -
  E_{\rm min},
  \label{eq:energy_change}
\end{equation}
where $\mathbf{m}_{\rm min}$ corresponds to the initial stable state. 
Strictly speaking, the exact solution of the LLG equation is necessary to evaluate 
the threshold current density from Eq. (\ref{eq:energy_change}). 
However, the LLG equation is a nonlinear equation, and it is difficult to obtain the exact solution. 
Instead, we approximate Eq. (\ref{eq:energy_change}) as 
\begin{equation}
  \int_{\mathbf{m}_{\rm d \pm}}^{\mathbf{m}_{\rm d}} dt 
  \frac{dE}{dt}
  \simeq 
  E_{\rm saddle}
  -
  E_{\rm min},
  \label{eq:approximated_energy_change}
\end{equation}
where $\mathbf{m}_{\rm d \pm}$ are the points on the constant energy curve of $E_{\rm saddle}$ 
and are located in the $xz$ plane; see Fig. \ref{fig:fig3}. 
We note that Eq. (\ref{eq:energy_change}) is well approximated by Eq. (\ref{eq:approximated_energy_change}) 
when $\mathbf{m}_{\rm d \pm}$ locate close to $\mathbf{m}_{\rm min}$, 
which means that $H_{\rm appl}/(4\pi M) \ll 1$. 
Note that the left hand side of Eq. (\ref{eq:approximated_energy_change}) can be evaluated 
in a similar manner to calculating Eq. (\ref{eq:balance_current}) 
because the integral range is on the constant energy curve. 
However, the integral range is limited to $[\mathbf{m}_{\rm d\pm},\mathbf{m}_{\rm d}]$ in Eq. (\ref{eq:approximated_energy_change}), 
while the range is over a periodic precession in Eq. (\ref{eq:balance_current}). 
The values of the integrals for these different regions are, in general, different. 
Since the value of the integral in Eq. (\ref{eq:approximated_energy_change}) is determined by the energy landscape, 
and the time-dependent solution of Eq. (\ref{eq:LLG}) is unnecessary, 
the integral in Eq. (\ref{eq:approximated_energy_change}) is more easily evaluated than that in Eq. (\ref{eq:energy_change}) [\onlinecite{bertotti09}]. 


The current density satisfying Eq. (\ref{eq:approximated_energy_change}) is given by 
\begin{equation}
\begin{split}
  j_{\rm th \pm}
  =&
  \frac{2 \alpha eMd}{\hbar \eta}
  \frac{\int_{\mathbf{m}_{\rm d \pm}}^{\mathbf{m}_{\rm d}} dt [\mathbf{H}^{2}-(\mathbf{m}\cdot\mathbf{H})^{2}]}
    {\int_{\mathbf{m}_{\rm d \pm}}^{\mathbf{m}_{\rm d}} dt [\mathbf{p}\cdot\mathbf{H}-(\mathbf{m}\cdot\mathbf{p})(\mathbf{m}\cdot\mathbf{H})]}
\\
  &+
  \frac{2ed}{\gamma \hbar \eta}
  \frac{E_{\rm saddle}-E_{\rm min}}{\int_{\mathbf{m}_{\rm d \pm}}^{\mathbf{m}_{\rm d}} dt [\mathbf{p}\cdot\mathbf{H}-(\mathbf{m}\cdot\mathbf{p})(\mathbf{m}\cdot\mathbf{H})]}.
  \label{eq:current_switch}
\end{split}
\end{equation}
Equation (\ref{eq:current_switch}) is the theoretical formula of the threshold current density 
and is the main result in this paper. 
This equation provides the estimation of the threshold current density with high accuracy. 
For example, the values of $j_{\rm th \pm}$ with the parameters used in Fig. \ref{fig:fig2} are 
$j_{\rm th +}=-7.7 \times 10^{6}$A/cm${}^{2}$ and $j_{\rm th -}=7.6 \times 10^{6}$A/cm${}^{2}$, 
which show good agreement with the numerical results in Fig. \ref{fig:fig2}. 
These values are estimated for $\theta_{H}=5^{\circ}$. 
Below, we show that the agreement between Eq. (\ref{eq:current_switch}) and the numerical simulation is obtained 
also for different values of $\theta_{H}$; see Fig. \ref{fig:fig5}. 
Note that $|j_{\rm th+}| \neq |j_{\rm th-}|$ because the magnetic field pointing in the positive $z$ direction breaks the symmetry 
between the magnetization dynamics moving to the positive and negative $z$ directions, 
although the difference is small. 
We emphasize that Eq. (\ref{eq:current_switch}) consists of two parts. 
One is proportional to $\alpha$ because this term arises from the energy dissipation due to the damping. 
The other is, on the other hand, independent of $\alpha$ but proportional to the energy barrier $E_{\rm saddle}-E_{\rm min}$. 


Equation (\ref{eq:current_switch}) can be simplified into a different form for $\theta_{H}=90^{\circ}$ 
(see also Appendix \ref{sec:AppendixC}). 
In this case, $E_{\rm saddle}=MH_{\rm appl}$, $E_{\rm min}=-MH_{\rm appl}$, and 
$\mathbf{m}_{\rm d\pm}=(\sqrt{1-z_{\rm d\pm}^{2}},0,z_{\rm d \pm})$ 
with $z_{\rm d \pm}=\pm 2 \sqrt{h(1-h)}$ and $h=H_{\rm appl}/(4\pi M)$. 
Then, we find that 
\begin{equation}
  \int_{\mathbf{m}_{\rm d \pm}}^{\mathbf{m}_{\rm d}} 
  dt 
  \left[
    \mathbf{p}
    \cdot
    \mathbf{H}
    -
    \left(
      \mathbf{m}
      \cdot
      \mathbf{p}
    \right)
    \left(
      \mathbf{m}
      \cdot
      \mathbf{H}
    \right)
  \right]
  =
  \mp
  \frac{\pi}{\gamma}
  (1-h)^{2},
  \label{eq:Melnikov_s_90deg}
\end{equation}
\begin{equation}
\begin{split}
 &
  \int_{\mathbf{m}_{\rm d \pm}}^{\mathbf{m}_{\rm d}} 
  dt 
  \left[
    \mathbf{H}^{2}
    -
    \left(
      \mathbf{m}
      \cdot
      \mathbf{H}
    \right)^{2}
  \right]
 \\
  &=
  \frac{16 \pi M}{3 \gamma}
  \sqrt{h(1-h)}
  \left(
    3
    -
    5h
    +
    2h^{2}
  \right).
  \label{eq:Melnikov_alpha_90deg}
\end{split}
\end{equation}
Therefore, Eq. (\ref{eq:current_switch}) becomes 
\begin{equation}
\begin{split}
  j_{\rm th \pm}(\theta_{H}=90^{\circ})
  =&
  \mp
  \frac{2 e M^{2}d}{\hbar\eta}
\\
  & \times 
  \left[
    \frac{16 \alpha }{3}
    \sqrt{
      \frac{h}{1-h}
    }
    \left(
      3
      -
      2h
    \right)
    +
    \frac{8h}{(1-h)^{2}}
  \right].
  \label{eq:threshold_current_in_plane}
\end{split}
\end{equation}
Note that $j_{\rm th \pm} \to 0$ in the limit of $h=H_{\rm appl}/(4\pi M) \to 0$, 
indicating that infinitesimal current can destabilize the stable state in the absence of the applied field. 


We note that both the positive and negative currents can destabilize the stable state in our picture, 
contrary to $j_{\rm c}$ having a fixed sign (positive for $\theta_{H}<90^{\circ}$). 
The physical meaning of this difference is as follows. 
Since the damping torque always dissipates energy from the free layer, 
positive energy should be supplied from the work done by spin torque to destabilize the stable state. 
In the derivation of $j_{\rm c}$, a steady precession around the stable state is assumed. 
On the precession trajectory, the spin torque has a component antiparallel to the damping torque when $m_{z} \lesssim 0$ 
and has a component parallel to the damping torque when $m_{z} \gtrsim 0$, for a positive current. 
The spin torque supplies energy to the free layer in the former case, 
but dissipates energy from the free layer in the latter case. 
Note that the trajectory slightly shifts to the positive direction due to the magnetic field having the positive $z$ component, 
i.e., the trajectory is not symmetric with respect to the $xy$ plane. 
Then, the work done by spin torque during the precession becomes finite and positive. 
The spin torque overcomes the damping torque when the current density becomes larger than $j_{\rm c}$. 
When the current direction is reversed, the work done by spin torque becomes negative, 
and thus, the spin torque cannot overcome the damping. 
As a result, the sign of $j_{\rm c}$ is positive. 
However, as emphasized above, a periodic precession around the easy axis assumed in the derivation of $j_{\rm c}$ is not excited in the present case. 
Instead, we focused on the magnetization dynamics from $\mathbf{m}_{\rm d\pm}$ to $\mathbf{m}_{\rm d}$. 
The work done by spin torque during $[\mathbf{m}_{\rm d-},\mathbf{m}_{\rm d}]$ becomes positive 
when the current has the positive sign. 
Similarly, the work during $[\mathbf{m}_{\rm d+},\mathbf{m}_{\rm d}]$ is positive 
when the current sign is negative. 
Therefore, both the positive and negative currents can destabilize the stable state by compensating the damping torque. 
Note also that the magnetization crosses the constant energy curve of $E_{\rm saddle}$ during a time shorter than a precession period around the $x$ axis. 
Therefore, an in-plane self-oscillation on a constant energy curve of $E_{\rm min} \le E \le E_{\rm saddle}$ around the $x$ axis cannot be excited in the present case. 


\subsection{Applicability of the present theory}
\label{sec:Applicability of the present theory}

There are two characteristic current scales, $j_{\rm c}$ and $j_{\rm th \pm}$, 
related to the magnetization dynamics, as discussed above. 
These two currents are defined from different mechanisms of the instability of the stable state. 
The instability condition of a precession around the stable state gives $j_{\rm c}$. 
On the other hand, $j_{\rm th \pm}$ was derived by the condition that 
the energy gain by the spin torque during a time shorter than the precession period becomes larger than the energy barrier 
between the stable state and the saddle point. 
The initial state is destabilized when the current magnitude becomes larger than ${\rm min}[j_{\rm c},j_{\rm th\pm}]$. 
For the present parameters, 
$j_{\rm th \pm}$ is smaller than $j_{\rm c}$, 
and therefore, $j_{\rm th \pm}$ determines the instability threshold. 
The condition that $j_{\rm th\pm}$ works well to estimate the instability of the stable state can be expressed as 
\begin{equation}
  \frac{j_{\rm th \pm}}{j_{\rm c}}
  <
  1.
  \label{eq:jc_condition_1}
\end{equation}
This is another important equation in this paper, 
guaranteeing the validity of our approach. 
Whether Eq. (\ref{eq:jc_condition_1}) is satisfied or not depends on 
the material parameters, as well as the applied field magnitude and angle. 
If Eq. (\ref{eq:jc_condition_1}) is unsatisfied, 
$j_{\rm c}$ determines the instability threshold, 
the magnetization moves to the out-of-plane region after the magnetization precesses around the in-plane axis. 


Note that the first term on the right hand side of Eq. (\ref{eq:current_switch}) is proportional to the damping constant $\alpha$, 
while the second term is independent of $\alpha$. 
On the other hand, Eq. (\ref{eq:jc}) is proportional to $\alpha$. 
Therefore, Eq. (\ref{eq:jc_condition_1}) is not satisfied when $\alpha$ becomes sufficiently small. 
When Eq. (\ref{eq:jc_condition_1}) is unsatisfied, 
$j_{\rm c}$ determines the instability of the stable state. 
Then, we can discuss the minimum value of $\alpha$ guaranteeing the applicability of Eq. (\ref{eq:jc_condition_1}) [\onlinecite{taniguchi13PRB}]. 
The value $\alpha$ which falls off from the condition in Eq. (\ref{eq:jc_condition_1}) 
for the parameters used in Fig. \ref{fig:fig2} is $\alpha< 1.7 \times 10^{-4}$. 
This value of $\alpha$ is at least one to two orders of magnitude smaller than 
the experimentally reported values for conventional ferromagnets used in spin-torque oscillator, such as CoFeB [\onlinecite{oogane06,tsunegi14}]. 
Therefore, we consider that $j_{\rm th \pm}$ determines the instability of the stable state for typical experiments. 


\subsection{In the presence of the angular dependence of the spin torque}

When the applied field points to the in-plane direction, $\theta_{H}=90^{\circ}$, 
the stable state corresponds to $\theta_{0}=90^{\circ}$, 
and the critical current density in Eq. (\ref{eq:jc}) diverges. 
This is because the work done by spin torque during the precession around the stable state becomes zero. 
Therefore, Eq. (\ref{eq:jc_condition_1}) is always satisfied for $\theta_{H}=90^{\circ}$. 

The divergence of $j_{\rm c}$ appears at a different field angle 
when the angular dependence of the spin torque is taken into account, 
although this term is neglected in the above calculation, for simplicity. 
In this case, Eq. (\ref{eq:H_s}) is replaced by 
\begin{equation}
  H_{\rm s}
  =
  \frac{\hbar \eta j}{2e(1+\lambda \mathbf{m}\cdot\mathbf{p})Md}.
  \label{eq:H_s_lambda}
\end{equation}
Then, the critical current density becomes 
\begin{equation}
  j_{\rm c}
  =
  \frac{2 \alpha eMd}{\hbar \eta P(\theta_{0})}
  \left[
    H_{\rm appl}
    \cos(\theta_{H}-\theta_{0})
    -
    4\pi M 
    \frac{\cos^{2}\theta_{0} + \cos 2 \theta_{0}}{2}
  \right],
  \label{eq:jc_lambda_system}
\end{equation}
where $P(\theta_{0})$ is given by 
\begin{equation}
  P(\theta_{0})
  =
  \frac{\cos\theta_{0}}{1+\lambda \cos\theta_{0}}
  +
  \frac{\lambda \sin^{2}\theta_{0}}{2(1+\lambda \cos\theta_{0})^{2}},
\end{equation}
see Appendix \ref{sec:AppendixA}. 
The divergence of the critical current density, Eq. (\ref{eq:jc_lambda_system}), occurs at the angle $\theta_{0}$ satisfying $P(\theta_{0})=0$. 
In particular, when $\theta_{H}=90^{\circ}$, 
Eqs. (\ref{eq:jc_lambda_system}) becomes 
\begin{equation}
  j_{\rm c}(\theta_{H}=90^{\circ})
  =
  \frac{4\alpha eMd}{\hbar \eta \lambda}
  \left(
    H_{\rm appl}
    +
    2\pi M
  \right).
  \label{eq:jc_lambda_in_plane}
\end{equation}
On the other hand, Eq. (\ref{eq:current_switch}) is generalized for finite $\lambda$ as 
\begin{equation}
\begin{split}
  j_{\rm th \pm}
  =&
  \frac{2 \alpha eMd}{\hbar \eta}
  \frac{\int_{\mathbf{m}_{\rm d \pm}}^{\mathbf{m}_{\rm d}} dt [\mathbf{H}^{2}-(\mathbf{m}\cdot\mathbf{H})^{2}]}
    {\int_{\mathbf{m}_{\rm d \pm}}^{\mathbf{m}_{\rm d}} dt [\mathbf{p}\cdot\mathbf{H}-(\mathbf{m}\cdot\mathbf{p})(\mathbf{m}\cdot\mathbf{H})]/(1+\lambda\mathbf{m}\cdot\mathbf{p})}
\\
  &+
  \frac{2ed}{\gamma \hbar \eta}
  \frac{E_{\rm saddle}-E_{\rm min}}{\int_{\mathbf{m}_{\rm d \pm}}^{\mathbf{m}_{\rm d}} dt [\mathbf{p}\cdot\mathbf{H}-(\mathbf{m}\cdot\mathbf{p})(\mathbf{m}\cdot\mathbf{H})]/(1+\lambda\mathbf{m}\cdot\mathbf{p})}.
  \label{eq:current_switch_lambda}
\end{split}
\end{equation}
Equation (\ref{eq:current_switch_lambda}) for $\theta_{H}=90^{\circ}$ is 
\begin{equation}
  j_{\rm th \pm}(\theta_{H}=90^{\circ})
  =
  \mp
  \frac{2 e Md}{\hbar \eta}
  4\pi M 
  \frac{\mathcal{N}}{\mathcal{D}_{\pm}},
  \label{eq:current_switch_lambda_in_plane}
\end{equation}
where $\mathcal{N}$ and $\mathcal{D}_{\pm}$ are 
\begin{equation}
\begin{split}
  \mathcal{N}
  =&
  4 \lambda^{2}
  \left[
    2 \alpha 
    (3-2h)
    (1-h)
    \sqrt{h(1-h)} 
    + 
    3h
  \right] 
\\
  & 
  \times
  \sqrt{1-4 \lambda^{2} h(1-h)}
\end{split}
\end{equation}
\begin{equation}
\begin{split}
  \mathcal{D}_{\pm}
  =&
  3 
  \left\{ 
    \sqrt{1-4\lambda^{2} h(1-h)}
    \left[
      \pi 
      \mp 
      4 \lambda \sqrt{h(1-h)}
    \right] 
  \right.
\\
  &
  \left.
    - 2 
    \left[
      1
      -
      2 \lambda^{2}(1-h)
    \right]
    \cos^{-1}
    \left[ 
      \pm 2 
      \lambda \sqrt{h(1-h)}
    \right]
  \right\},
\end{split}
\end{equation}
see Appendix \ref{sec:AppendixB}. 
In the presence of a finite $\lambda$, $|j_{\rm th +}| \neq |j_{\rm th -}|$ even for $\theta_{H}=90^{\circ}$. 
Eq. (\ref{eq:current_switch_lambda_in_plane}) reproduces Eq. (\ref{eq:threshold_current_in_plane}) in the limit of $\lambda \to 0$. 
The currents, $j_{\rm c}$ and $j_{\rm th \pm}$, in Eq. (\ref{eq:jc_condition_1}) should be replaced by 
Eqs. (\ref{eq:jc_lambda_system}) and (\ref{eq:current_switch_lambda}) in the presence of the angular dependence of the spin torque.

\subsection{Validity of Eq. (\ref{eq:current_switch}) and condition to excite out-of-plane self-oscillation}

In this section, we confirm the validity of Eq. (\ref{eq:current_switch}) for a wide range of $\theta_{H}$ 
by comparing with the numerical simulation of the LLG equation. 

\begin{figure}
\centerline{\includegraphics[width=1.0\columnwidth]{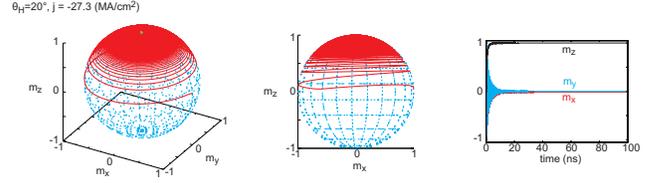}}
\caption{
         The trajectories of the magnetization dynamics on the unit spheres and 
         the time evolutions of the magnetization components for $\theta_{H}=20^{\circ}$ and $j=-27.3 \times 10^{6}$ A/cm${}^{2}$. 
         \vspace{-3ex}}
\label{fig:fig4}
\end{figure}

Before the comparison, we briefly discuss the definition of the threshold current density estimated from the numerical simulation. 
We emphasize that $j_{\rm th \pm}$ just determines the instability of the stable state, 
and does not guarantee the existence of the out-of-plane self-oscillation. 
The out-of-plane self-oscillation is excited when a condition, 
\begin{equation}
  \frac{j(E)}{j_{\rm th \pm}}
  >
  1,
  \label{eq:condition_oscillation_2}
\end{equation}
is satisfied [\onlinecite{taniguchi15}], 
where the range of $E$ is $E_{\rm saddle}<E \le E_{\rm max \pm}$. 
Note that the reason why the out-of-plane self-oscillations appear in Figs. \ref{fig:fig2}(b) and \ref{fig:fig2}(d) is that 
there exists a certain $E$ satisfying Eq. (\ref{eq:condition_oscillation_2}). 
On the other hand, when Eq. (\ref{eq:condition_oscillation_2}) is not satisfied for any value of $E$, 
the magnetization moves to the point close to $-(+)\mathbf{e}_{z}$ for a positive (negative) current above $j_{\rm th \pm}$ 
because the spin-torque magnitude becomes sufficiently strong, 
and the magnetization eventually becomes parallel or antiparallel to the magnetization of the pinned layer, $\mathbf{p}=+\mathbf{e}_{z}$. 
Figure \ref{fig:fig4} shows an example of such dynamics, 
where $\theta_{H}=20^{\circ}$ and the current density is close to the threshold value, $-27.3 \times 10^{6}$A/cm${}^{2}$, for this $\theta_{H}$. 
As shown, the magnetization finally becomes almost parallel to the $z$ axis. 
Such magnetization dynamics was observed experimentally [\onlinecite{houssameddine07}]. 
The threshold current density evaluated from the numerical simulation 
should be defined as the current density above which  
the magnetization shows a stable out-of-plane self-oscillation or the magnetization moves to the points close to $\pm\mathbf{e}_{z}$. 
The detail of the method numerically defining the threshold current density is summarized in Appendix \ref{sec:AppendixD}. 




\begin{figure*}
\centerline{\includegraphics[width=2.0\columnwidth]{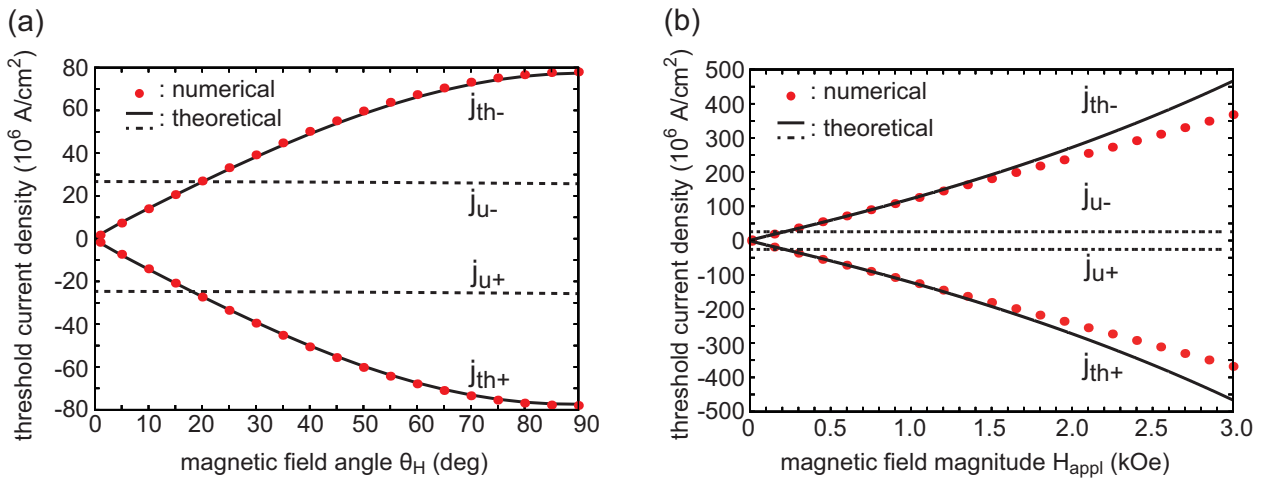}}
\caption{
         Dependences of the threshold current densities estimated by numerically solving the LLG equation (\ref{eq:LLG}) (dots), 
         the theoretical instability threshold $j_{\rm th \pm}$, Eq. (\ref{eq:current_switch}), (solid lines), 
         and the current densities $j_{\rm u \pm}$, Eq. (\ref{eq:j_max}), (dotted lines)
         on (a) the applied field angle $\theta_{H}$ ($H_{\rm appl}=650$ Oe) and (b) the magnitude $H_{\rm appl}$ ($\theta_{H}=90^{\circ}$). 
         \vspace{-3ex}}
\label{fig:fig5}
\end{figure*}



We study the validity of Eq. (\ref{eq:current_switch}) by comparing with the threshold current estimated by numerically solving Eq. (\ref{eq:LLG}) 
for several values of $\theta_{H}$ and $H_{\rm appl}$. 
The threshold current density estimated from the numerical simulation of the LLG equation is shown by dots in Fig. \ref{fig:fig5}(a), 
where the magnetic field angle $\theta_{H}$ varies in the range of $0 < \theta_{H} \le 90^{\circ}$ 
while the magnitude $H_{\rm appl}$ is fixed to 650 Oe. 
We also shows the value of $j_{\rm th\pm}$ evaluated from Eq. (\ref{eq:current_switch}) by solid lines. 
We find a good agreement between the numerical and theoretical results, 
supporting the validity of Eq. (\ref{eq:current_switch}). 
The comparison between the numerically evaluated instability threshold and the analytical formula, Eq. (\ref{eq:threshold_current_in_plane}), 
for several values of the field magnitude $H_{\rm appl}$ is shown in Fig. \ref{fig:fig5}(b), 
where the field angle is fixed to $\theta_{H}=90^{\circ}$. 
The theoretical formula agrees with the numerical result when $H_{\rm appl}/(4\pi M) \ll 1$, 
while the numerical result becomes different with the theoretical formula for relatively large magnetic field. 
This is because the derivation of the theoretical formula, Eq. (\ref{eq:current_switch}), assumes that $H_{\rm appl}/(4\pi M) \ll 1$, 
as mentioned below Eq. (\ref{eq:approximated_energy_change}). 
The current magnitude above which the difference between the theoretical and numerical results appears is on the order of $10^{8}$ A/cm${}^{2}$, 
which is one to two orders of magnitude larger than the current magnitude used in typical experiments [\onlinecite{houssameddine07,suto12,bosu16,hiramatsu16}]. 
Therefore, we consider that the present formula works well to analyze experiments for wide range of the applied field angle and magnitude. 



We notice that $j(E)$ is an increasing function of $E$ for the out-of-plane self-oscillation when $\lambda=0$, as in the zero field case [\onlinecite{lee05,silva10}]. 
Then, there is a certain $E$ satisfying Eq. (\ref{eq:condition_oscillation_2}) if 
\begin{equation}
  \frac{j_{\rm u \pm}}{j_{\rm th \pm}}
  >
  1, 
  \label{eq:condition_oscillation_3}
\end{equation}
is satisfied, 
where $j_{\rm u\pm}$ are Eq. (\ref{eq:balance_current}) at the unstable states, $E=E_{\rm max \pm}$, 
\begin{equation}
  j_{\rm u \pm}
  \equiv 
  \lim_{E \to E_{\rm max \pm}}j(E).
  \label{eq:j_max}
\end{equation}
The dependences of $j_{\rm th \pm}$ and $j_{\rm u \pm}$ on the field angle $\theta_{H}$ and the magnitude $H_{\rm appl}$ 
are also shown in Figs. \ref{fig:fig5}(a) and \ref{fig:fig5}(b), respectively. 
It is shown that $j_{\rm u \pm}$ is almost independent of $\theta_{H}$ and $H_{\rm appl}$, 
while $j_{\rm th \pm}$ increases with increasing these parameters. 
For example, we find that $j_{\rm u \pm}/j_{\rm th \pm}>1$ for $\theta_{H} < 20^{\circ}$. 
This result indicates that the out-of-plane self-oscillation can be excited for $\theta_{H} < 20^{\circ}$ for the present parameters. 
This finding is consistent with the numerical results shown in Figs. \ref{fig:fig2} and \ref{fig:fig4}, 
supporting the validity of our argument. 
We notice that the linearized LLG equation is useful to estimate $\lim_{E \to E_{\rm max \pm}}j(E)$ 
by replacing $(\theta_{0},\varphi_{0})$ with the zenith angle corresponding to the maximum point; see Eq. (\ref{eq:jc_general}). 
In particular, when $\theta_{H}=90^{\circ}$, 
the unstable states locate at $\mathbf{m}_{\rm u+}=(-H_{\rm appl}/(4\pi M),0,\sqrt{1-[H_{\rm appl}/(4\pi M)]^{2}})$ 
and $\mathbf{m}_{\rm u-}=(-H_{\rm appl}/(4\pi M),0,-\sqrt{1-[H_{\rm appl}/(4\pi M)]^{2}})$. 
Then, we find that (see also Appendix \ref{sec:AppendixB}) 
\begin{equation}
\begin{split}
  j_{\rm u \pm}(\theta_{H}=90^{\circ})
  &=
  \mp
  \frac{2\alpha eMd}{\hbar \eta \sqrt{1-h^{2}}}
  4\pi M 
  \left(
    1
    -
    \frac{h^{2}}{2}
  \right).
  \label{eq:jc_max}
\end{split}
\end{equation}
This equation indicates that $j_{\rm u \pm}(\theta_{H}=90^{\circ}) \simeq \mp [2 \alpha eMd/(\hbar \eta)] 4\pi M$ for $h \ll 1$, 
i.e., $j_{\rm u \pm}$ is almost independent of $H_{\rm appl}$, 
which is consistent with the result shown in Fig. \ref{fig:fig5}(b). 






\section{Conclusion}
\label{sec:Conclusion}

In conclusion, we studied the theoretical conditions to excite the self-oscillation 
in a spin-torque oscillator consisting of an in-plane magnetized free layer and a perpendicularly magnetized pinned layer 
in the presence of an external magnetic field pointing in an arbitrary direction. 
The numerical simulation in Fig. \ref{fig:fig2} showed that the initial stable state is destabilized by current density 
much smaller than the critical current density estimated from the linearized LLG equation, Eq. (\ref{eq:jc}). 
The fact implies that the linearized LLG equation is no longer applicable to evaluate the instability threshold in the present system. 
Then, we derived the theoretical formula of the threshold current density, Eq. (\ref{eq:current_switch}), 
by focusing on the transition of the magnetization from the stable state to the out-of-plane precession 
during a time shorter than a precession period around the stable state. 
The derived formula consists of two parts, 
where one is proportional to the damping constant $\alpha$, 
while the other is independent of $\alpha$ but proportional to the energy barrier $E_{\rm saddle}-E_{\rm min}$ for the transition. 
A good agreement between the numerical simulation and our formula, Eq. (\ref{eq:current_switch}), is obtained in Fig. \ref{fig:fig5}, 
indicating the validity of the formula. 
The condition that our formula of the threshold current density works better than the linear analysis to instability threshold is Eq. (\ref{eq:jc_condition_1}). 
We also derived the theoretical condition, Eq. (\ref{eq:condition_oscillation_2}), to stabilize the out-of-plane self-oscillation. 


\section*{Acknowledgement}

The authors express gratitude to 
Shinji Yuasa, Kay Yakushiji, Akio Fukushima, Shingo Tamaru, Sumito Tsunegi, Ryo Hiramatsu, Yoichi Shiota, Takehiko Yorozu, Hirofumi Suto, and Kiwamu Kudo 
for valuable discussion they had with us.  
This work is supported by Japan Society and Technology Agency (JST) strategic innovation promotion program 
"Development of new technologies for 3D magnetic recording architecture". 



\appendix


\section{Derivation of linearized LLG equation}
\label{sec:AppendixA}

In this Appendix, we show the detail of the derivation of Eq. (\ref{eq:linearized_LLG}). 
For generality, we consider a ferromagnet having uniaxial anisotropies along the $x$, $y$, and $z$ axes 
with an external magnetic field applied in an arbitrary direction. 
The magnetic field is given by 
\begin{equation}
  \mathbf{H}
  =
  \begin{pmatrix}
    H_{\rm appl} \sin\theta_{H} \cos\varphi_{H} - 4\pi M \tilde{N}_{x} m_{x} \\
    H_{\rm appl} \sin\theta_{H} \sin\varphi_{H} - 4\pi M \tilde{N}_{y} m_{y} \\
    H_{\rm appl} \cos\theta_{H} - 4\pi M \tilde{N}_{z} m_{z}
  \end{pmatrix}.
  \label{eq:field_general}
\end{equation}
The generalized demagnetization coefficient $\tilde{N}_{i}$ ($i=x,y,z$) is defined as 
$4\pi M \tilde{N}_{i}=4\pi M N_{i} - H_{{\rm K}i}$, 
where $4\pi M N_{i}$ is the shape anisotropy (demagnetization) field with $N_{x}+N_{y}+N_{z}=1$, 
while $H_{{\rm K}i}$ is the crystalline or interface anisotropy field. 
The energy density $E=-M \int d \mathbf{m}\cdot\mathbf{H}$ is 
\begin{equation}
\begin{split}
  \frac{E}{M}
  =&
  -H_{\rm appl}
  \left[
    \sin\theta_{H}
    \sin\theta
    \cos(\varphi_{H}-\varphi)
    +
    \cos\theta_{H}
    \cos\theta
  \right]
\\
  &+
  2\pi M 
  \tilde{N}_{x}
  \sin^{2}\theta
  \cos^{2}\varphi
  +
  2\pi M
  \tilde{N}_{y}
  \sin^{2}\theta
  \sin^{2}\varphi
\\
  &+
  2\pi M 
  \tilde{N}_{z}
  \cos^{2}\theta.
  \label{eq:energy_general}
\end{split}
\end{equation}
The system in the main text corresponds to the case of 
$\tilde{N}_{x}=\tilde{N}_{y}=0$, $\tilde{N}_{z}=1$, and $\varphi_{H}=0$. 


Since we are interested in a small oscillation of the magnetization around the stable state, 
the zenith and azimuth angles corresponding to the stable state should be identified. 
The stable state is determined by the conditions that $\partial E/\partial \theta = \partial E/\partial \varphi=0$, 
which are explicitly given by 
\begin{equation}
\begin{split}
  &
  H_{\rm appl}
  \left[
   \sin\theta_{H}
   \cos\theta
   \cos(\varphi_{H}-\varphi)
   -
   \cos\theta_{H}
   \sin\theta
  \right]
\\
  &-
  4\pi M 
  \tilde{N}_{x}
  \sin\theta
  \cos\theta
  \cos^{2}\varphi
  -
  4\pi M 
  \tilde{N}_{y}
  \sin\theta
  \cos\theta
  \sin^{2}\varphi
\\
  &+
  4\pi M 
  \tilde{N}_{z}
  \sin\theta
  \cos\theta
  =
  0,
  \label{eq:stable_theta}
\end{split}
\end{equation}
\begin{equation}
\begin{split}
  &
  H_{\rm appl}
  \sin\theta_{H}
  \sin\theta
  \sin(\varphi_{H}-\varphi)
\\
  &+
  4\pi M 
  \tilde{N}_{x}
  \sin^{2}\theta
  \sin\varphi
  \cos\varphi
  -
  4\pi M 
  \tilde{N}_{y}
  \sin^{2}\theta
  \sin\varphi
  \cos\varphi
  =
  0.
  \label{eq:stable_varphi}
\end{split}
\end{equation}
Let us denote the zenith and azimuth angles satisfying Eqs. (\ref{eq:stable_theta}) and (\ref{eq:stable_varphi}) as $(\theta_{0},\varphi_{0})$. 
As mentioned in the main text, 
we introduce the $XYZ$ coordinate where the $Z$ axis is parallel to the stable state $(\theta_{0},\varphi_{0})$. 
The rotation to the $xyz$ coordinate to the $XYZ$ coordinate is described by the rotation matrix 
\begin{equation}
  \mathsf{R}
  =
  \begin{pmatrix}
    \cos\theta_{0} & 0 & -\sin\theta_{0} \\
    0 & 1 & 0 \\
    \sin\theta_{0} & 0 & \cos\theta_{0}
  \end{pmatrix}
  \begin{pmatrix}
    \cos\varphi_{0} & \sin\varphi_{0} & 0 \\
    -\sin\varphi_{0} & \cos\varphi_{0} & 0 \\
    0 & 0 & 1
  \end{pmatrix}.
\end{equation}
The relations between the components of $\mathbf{m}$ in the $xyz$ and $XYZ$ coordinates are 
$m_{x}=m_{X}\cos\theta_{0}\cos\varphi_{0}-m_{Y}\sin\varphi_{0}+m_{Z}\sin\theta_{0}\cos\varphi_{0}$, 
$m_{y}=m_{X}\cos\theta_{0}\sin\varphi_{0}+m_{Y}\cos\varphi_{0}+m_{Z}\sin\theta_{0}\sin\varphi_{0}$, 
$m_{z}=-m_{X}\sin\theta_{0}+m_{Z}\cos\theta_{0}$. 
Also, the magnetic field in the $XYZ$ coordinate is 
\begin{equation}
  \mathbf{H}
  =
  \begin{pmatrix}
    H_{XX}m_{X} + H_{XY}m_{Y} \\
    H_{YX}m_{X} + H_{YY}m_{Y} \\
    H_{ZX}m_{X} + H_{ZY}m_{Y} + H_{ZZ}
  \end{pmatrix},
\end{equation}
where
\begin{equation}
\begin{split}
  H_{XX}
  =&
  -4\pi M 
  \tilde{N}_{x}
  \cos^{2}\theta_{0}
  \cos^{2}\varphi_{0}
  -
  4\pi M
  \tilde{N}_{y}
  \cos^{2}\theta_{0}
  \sin^{2}\varphi_{0}
\\
  &
  - 4\pi M 
  \tilde{N}_{z}
  \sin^{2}\theta_{0},
\end{split}
\end{equation}
\begin{equation}
  H_{XY}
  =
  H_{YX}
  =
  -4\pi M 
  \left(
    \tilde{N}_{y}
    -
    \tilde{N}_{x}
  \right)
  \cos\theta_{0}
  \sin\varphi_{0}
  \cos\varphi_{0},
\end{equation}
\begin{equation}
  H_{YY}
  =
  -4\pi M 
  \tilde{N}_{x}
  \sin^{2}\varphi_{0}
  -
  4\pi M 
  \tilde{N}_{y}
  \cos^{2}\varphi_{0},
\end{equation}
\begin{equation}
\begin{split}
  H_{ZX}
  =&
  -4\pi M 
  \tilde{N}_{x}
  \sin\theta_{0}
  \cos\theta_{0}
  \cos^{2}\varphi_{0}
\\
  &-
  4\pi M 
  \tilde{N}_{y}
  \sin\theta_{0}
  \cos\theta_{0}
  \sin^{2}\varphi_{0}
\\
  &+
  4\pi M 
  \tilde{N}_{z}
  \sin\theta_{0}
  \cos\theta_{0},
\end{split}
\end{equation}
\begin{equation}
  H_{ZY}
  =
  -4\pi M 
  \left(
    \tilde{N}_{y}
    -
    \tilde{N}_{x}
  \right)
  \sin\theta_{0}
  \sin\varphi_{0}
  \cos\varphi_{0},
\end{equation}
\begin{equation}
\begin{split}
  H_{ZZ}
  =&
  H_{\rm appl}
  \left[
    \sin\theta_{H}
    \sin\theta_{0}
    \cos(\varphi_{H}-\varphi_{0})
    +
    \cos\theta_{H}
    \cos\theta_{0}
  \right]
\\
  &-
  4\pi M
  \tilde{N}_{x}
  \sin^{2}\theta_{0}
  \cos^{2}\varphi_{0}
  -
  4\pi M
  \tilde{N}_{y}
  \sin^{2}\theta_{0}
  \sin^{2}\varphi_{0}
\\
  &
  - 
  4\pi M
  \tilde{N}_{z}
  \cos^{2}\theta_{0}. 
\end{split}
\end{equation}
Similarly, the magnetization of the pinned layer $\mathbf{p}=(p_{x},p_{y},p_{z})=(\sin\theta_{\rm p}\cos\varphi_{\rm p},\sin\theta_{\rm p}\sin\varphi_{\rm p},\cos\theta_{\rm p})$ in the $xyz$ coordinate 
transforms in the $XYZ$ coordinate to 
\begin{equation}
  \mathbf{p}
  \equiv
  \begin{pmatrix}
    p_{X} \\
    p_{Y} \\
    p_{Z}
  \end{pmatrix}
  =
  \begin{pmatrix}
    \sin\theta_{\rm p}\cos\theta_{0}\cos(\varphi_{\rm p}-\varphi_{0})-\cos\theta_{\rm p}\sin\theta_{0} \\
    \sin\theta_{\rm p}\sin(\varphi_{\rm p}-\varphi_{0}) \\
    \sin\theta_{\rm p}\sin\theta_{0}\cos(\varphi_{\rm p}-\varphi_{0})+\cos\theta_{\rm p}\cos\theta_{0}
  \end{pmatrix}.
\end{equation}


Now we consider a small oscillation of the magnetization around the stable state. 
Using the approximations $m_{Z} \simeq 1$ and $|m_{X}|,|m_{Y}| \ll 1$, 
the LLG equation is linearized as 
\begin{equation}
\begin{split}
  &
  \frac{1}{\gamma}
  \frac{d}{dt}
  \begin{pmatrix}
    m_{X} \\
    m_{Y}
  \end{pmatrix}
\\
  &+
  \begin{pmatrix}
    -H_{YX}-H_{\rm s}p_{Z}+\alpha H_{X} & H_{Y}-\alpha H_{XY} \\
    -H_{X}-\alpha H_{YX} & H_{XY}-H_{\rm s}p_{Z}+\alpha H_{Y}
  \end{pmatrix}
  \begin{pmatrix}
    m_{X} \\
    m_{Y}
  \end{pmatrix}
\\
  &=
  -H_{\rm s}
  \begin{pmatrix}
    p_{X} \\
    p_{Y}
  \end{pmatrix},
  \label{eq:LLG_linear_general}
\end{split}
\end{equation}
where $H_{X}=H_{ZZ}-H_{XX}$ and $H_{Y}=H_{ZZ}-H_{YY}$. 
The terms proportional to $\alpha H_{\rm s}$ are neglected because these terms are on the order of $\alpha^{2}$. 
The condition that the trace of the coefficient matrix is zero gives 
\begin{equation}
  j_{\rm c}
  =
  \frac{2\alpha eMd}{\hbar \eta p_{Z}}
  \left(
    \frac{H_{X} + H_{Y}}{2}
  \right).
  \label{eq:jc_general}
\end{equation}
Substituting $\tilde{N}_{x}=\tilde{N}_{y}=0$, $\tilde{N}_{z}=1$, $\varphi_{H}=0$, and $\theta_{\rm p}=0$, 
Eq. (\ref{eq:jc_general}) reproduces Eq. (\ref{eq:jc}). 
On the other hand, in the case of the in-plane magnetized system considered in Ref. [\onlinecite{grollier03}], 
i.e., $4\pi M \tilde{N}_{x}=-H_{\rm K}$, $\tilde{N}_{y}=0$, $\tilde{N}_{z}=1$, 
$\theta_{H}=90^{\circ}$, $\varphi_{H}=0$, $\theta_{\rm p}=90^{\circ}$, and $\varphi_{\rm p}=0$, 
we find that $H_{XX}=-4\pi M$, $H_{YY}=0$, and $H_{ZZ}=H_{\rm appl}+H_{\rm K}$, 
where $H_{\rm K}$ is the in-plane anisotropy. 
Then, the critical current density becomes $j_{\rm c}=[2\alpha eMd/(\hbar \eta)](H_{\rm appl}+H_{\rm K}+2\pi M)$, 
which is consistent with the result in Ref. [\onlinecite{grollier03}]. 


The angular dependence of the spin torque, characterized by the factor $1/(1+\lambda\mathbf{m}\cdot\mathbf{p})$, can be taken into account as follows. 
As mentioned in the main text, 
$H_{\rm s}$ in this case is given by Eq. (\ref{eq:H_s_lambda}). 
In this case, Eq. (\ref{eq:H_s}) is replaced by Eq. (\ref{eq:H_s_lambda}). 
The factor $1/(1+\lambda\mathbf{m}\cdot\mathbf{p})$ is linearized as 
\begin{equation}
\begin{split}
  \frac{1}{1+\lambda\mathbf{m}\cdot\mathbf{p}}
  &=
  \frac{1}{1+\lambda m_{Z}p_{Z}}
  \frac{1}{1+\frac{\lambda (m_{X}p_{X}+m_{Y}p_{Y})}{1+\lambda m_{Z}p_{Z}}}
\\
  &\simeq
  \frac{1}{1+\lambda p_{Z}}
  \left[
    1
    -
    \frac{\lambda(m_{X}p_{X}+m_{Y}p_{Y})}{1+\lambda p_{Z}}
  \right].
\end{split}
\end{equation}
We introduce the following notations,
\begin{equation}
  H_{\rm s}^{(0)}
  =
  \frac{\hbar \eta j}{2e(1+\lambda p_{Z})Md},
\end{equation}
\begin{equation}
  \Lambda
  =
  \frac{\lambda}{1+\lambda p_{Z}}.
\end{equation}
Then, Eq. (\ref{eq:LLG_linear_general}) becomes 
\begin{equation}
  \frac{1}{\gamma}
  \frac{d}{dt}
  \begin{pmatrix}
    m_{X} \\
    m_{Y} 
  \end{pmatrix}
  +
  \mathsf{M}
  \begin{pmatrix}
    m_{X} \\
    m_{Y}
  \end{pmatrix}
  =
  -H_{\rm s}^{(0)}
  \begin{pmatrix}
    p_{X} \\
    p_{Y}
  \end{pmatrix},
\end{equation}
where the components of the $2 \times 2$ matrix $\mathsf{M}$ are 
\begin{equation}
  \mathsf{M}_{1,1}
  =
  -H_{YX}
  -
  H_{\rm s}^{(0)}
  \left(
    p_{Z}
    +
    \Lambda p_{X}^{2}
  \right)
  +
  \alpha H_{X},
\end{equation}
\begin{equation}
  \mathsf{M}_{1,2}
  =
  H_{Y}
  -
  H_{\rm s}^{(0)}
  \Lambda
  p_{X}
  p_{Y}
  -
  \alpha H_{XY},
\end{equation}
\begin{equation}
  \mathsf{M}_{2,1}
  =
  -H_{X}
  -
  H_{\rm s}^{(0)}
  \Lambda
  p_{X}
  p_{Y}
  -
  \alpha H_{YX},
\end{equation}
\begin{equation}
  \mathsf{M}_{2,2}
  =
  H_{XY}
  -
  H_{\rm s}^{(0)}
  \left(
    p_{Z}
    +
    \Lambda
    p_{Y}^{2}
  \right)
  +
  \alpha H_{Y}.
\end{equation}
Then, the critical current determined by the condition ${\rm Tr}[\mathsf{M}]=0$ is 
\begin{equation}
  j_{\rm c}
  =
  \frac{2\alpha e (1+\lambda p_{Z})Md}{\hbar \eta [p_{Z}+\frac{\Lambda(1-p_{Z}^{2})}{2}]}
  \left(
    \frac{H_{X}+H_{Y}}{2}
  \right).
  \label{eq:jc_lambda}
\end{equation}
Equation (\ref{eq:jc_lambda_system}) is obtained from Eq. (\ref{eq:jc_lambda}) 
by substituting $\tilde{N}_{x}=\tilde{N}_{y}=0$, $\tilde{N}_{z}=1$, $\varphi_{H}=0$, and $\theta_{\rm p}=0$, 


\section{Derivations of Eqs. (\ref{eq:current_switch_lambda_in_plane}) and (\ref{eq:jc_max})}
\label{sec:AppendixB}

Let us show the derivation of Eq. (\ref{eq:jc_max}). 
As mentioned in the main text, $j_{\rm u \pm}$ can be obtained from the linearized LLG equation. 
Here, we show that $j_{\rm u \pm}$ can also be obtained as $j_{\rm u \pm}=\lim_{E \to E_{\rm max \pm}}j(E)$. 
This method provides an example of the calculation of Eq. (\ref{eq:balance_current}). 

\begin{figure}
\centerline{\includegraphics[width=1.0\columnwidth]{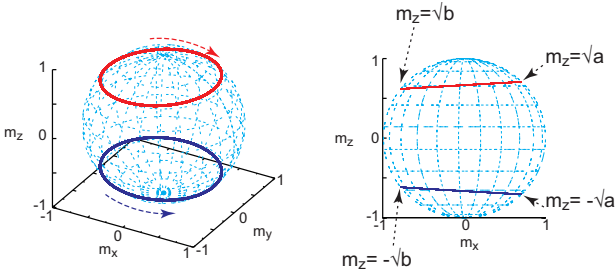}}
\caption{
         The examples of the out-of-plane precession trajectories (constant energy curves), 
         where the arrows indicate the precession directions. 
         The constant energy curves cross the $xz$ plane at $m_{z}=\pm\sqrt{a},\pm\sqrt{b}$. 
         \vspace{-3ex}}
\label{fig:fig6}
\end{figure}

Note that the maximum energies located at $\mathbf{m} \simeq \pm \mathbf{e}_{z}$, $E_{\rm max+}=E_{\rm max-}$, are identical for $\theta_{H}=90^{\circ}$, 
and the corresponding energy density is $E_{\rm max}=(4\pi M^{2}/2)(1+h^{2})$. 
Then, let us investigate $\lim_{E \to E_{\rm max}}j(E)$. 
Equation (\ref{eq:balance_current}) can be rewritten as 
\begin{equation}
  j(E)
  =
  \frac{2\alpha eMd}{\hbar \eta}
  \frac{\mathscr{N}_{\alpha}}{\mathscr{N}_{\rm s}},
  \label{eq:ju_in_plane_def}
\end{equation}
where $\mathscr{N}_{\rm s}$ and $\mathscr{N}_{\alpha}$ are, respectively, given by 
\begin{equation}
\begin{split}
  \mathscr{N}_{\rm s}
  &=
  \gamma
  \int 
  dt 
  \left[
    \mathbf{p}
    \cdot
    \mathbf{H}
    -
    \left(
      \mathbf{m}
      \cdot
      \mathbf{p}
    \right)
    \left(
      \mathbf{m}
      \cdot
      \mathbf{H}
    \right)
  \right]
\\
  &=
  -\frac{1}{h}
  \int 
  \frac{dm_{z}}{m_{y}}
  \left[
    m_{z}
    +
    \left(
      h m_{x}
      -
      m_{z}^{2}
    \right)
    m_{z}
  \right]
\\
  &=
  \int 
  dm_{z}
  \frac{m_{z}^{3}-2(1-\epsilon)m_{z}}{\sqrt{(a-m_{z}^{2})(m_{z}^{2}-b)}},
  \label{eq:integral_s}
\end{split}
\end{equation}
\begin{equation}
\begin{split}
  \mathscr{N}_{\alpha}
  &=
  \gamma
  \int 
  dt 
  \left[
    \mathbf{H}^{2}
    -
    \left(
      \mathbf{m}
      \cdot
      \mathbf{H}
    \right)^{2}
  \right]
\\
  &=
  (4\pi M)^{2}
  \gamma
  \int 
  dt 
  \left[
    h^{2}
    +
    m_{z}^{2}
    -
    \left(
      h m_{x}
      -
      m_{z}^{2}
    \right)^{2}
  \right]
\\
  &=
  -2\pi M
  \int 
  dm_{z}
  \frac{m_{z}^{4}-4(1-\epsilon)m_{z}^{2}+4(\epsilon^{2}-h^{2})}{\sqrt{(a-m_{z}^{2})(m_{z}^{2}-b)}}.
  \label{eq:integral_alpha}
\end{split}
\end{equation}
Here, we use the relation $dm_{z}/dt=\gamma H_{\rm appl} m_{y}$ obtained from 
the LLG equation on a constant energy curve, $d \mathbf{m}/dt=-\gamma \mathbf{m} \times \mathbf{H}$, with $\theta_{H}=90^{\circ}$ [\onlinecite{taniguchi14}]. 
The integral ranges of these integrals are discussed below. 
Equations (\ref{eq:Melnikov_s}) and (\ref{eq:Melnikov_alpha}) relate to Eqs. (\ref{eq:integral_s}) and (\ref{eq:integral_alpha}) via 
$\mathscr{W}_{\rm s}=2 MH_{\rm s} \mathscr{N}_{\rm s}$ and $\mathscr{W}_{\alpha}=-2 \alpha M \mathscr{N}_{\alpha}$, 
where the numerical factor $2$ appears by restricting the integral regions for $m_{y}>0$, according to the symmetry [\onlinecite{taniguchi14}]. 
The parameters $a$ and $b$ are given by 
\begin{equation}
  a
  =
  2 
  \left(
    \epsilon
    -
    h^{2}
    +
    h \sqrt{1+h^{2}-2\epsilon}
  \right),
\end{equation}
\begin{equation}
  b
  =
  2 
  \left(
    \epsilon
    -
    h^{2}
    -
    h \sqrt{1+h^{2}-2\epsilon}
  \right),
\end{equation}
where $\epsilon=E/(4 \pi M^{2})$ is the normalized energy density. 
The physical meanings of $a$ and $b$ are as follows. 
Figure \ref{fig:fig6} shows the examples of the out-of-plane precession trajectories (constant energy curves) in the regions of $m_{z}>0$ and $m_{z}<0$. 
The precession directions are indicated by the arrows. 
The constant energies curves cross the $xz$ plane at the points $m_{z}=\pm \sqrt{a},\pm \sqrt{b}$. 
When we focus on the out-of-plane precession for $m_{z}>0$, 
the integral ranges of Eqs. (\ref{eq:integral_s}) and (\ref{eq:integral_alpha}) are $\sqrt{b} \le m_{z} \le \sqrt{a}$. 
On the other hand, for the out-of-plane precession for $m_{z}<0$, 
the integral range is $-\sqrt{a} \le m_{z} \le -\sqrt{b}$. 
Below, we calculate Eqs. (\ref{eq:integral_s}) and (\ref{eq:integral_alpha}) for $m_{z}<0$. 
For $m_{z}>0$, the sign of $\mathscr{N}_{\rm s}$ is changed.



We notice that Eqs. (\ref{eq:integral_s}) and (\ref{eq:integral_alpha}) are expressed as 
$\mathscr{N}_{\rm s}=I_{3}-2 (1-\epsilon)I_{1}$ and 
$\mathscr{N}_{\alpha}=-2\pi M [I_{4} - 4(1-\epsilon) I_{2} + 4 (\epsilon^{2}-h^{2}) I_{0}]$, respectively, 
where $I_{n}$ ($n=0,1,2,3,4$) is 
\begin{equation}
\begin{split}
  I_{n}
  &=
  \int_{-\sqrt{a}}^{-\sqrt{b}} 
  dz 
  \frac{z^{n}}{\sqrt{(a-z^{2})(z^{2}-b)}}
\\
  &=
  \int_{0}^{1} 
  ds 
  \frac{(-\sqrt{a}\sqrt{1-k^{2}s^{2}})^{n}}{\sqrt{a} \sqrt{(1-s^{2})(1-k^{2}s^{2})}}.
\end{split}
\end{equation}
The modulus $k$ is 
\begin{equation}
  k
  =
  \sqrt{
    1
    -
    \frac{b}{a}
  }.
\end{equation}
The following formulas are useful to calculate $\mathscr{N}_{\rm s}$ and $\mathscr{N}_{\alpha}$; 
\begin{equation}
  I_{0}
  =
  \frac{1}{\sqrt{a}}
  \int_{0}^{1}
  \frac{ds}{\sqrt{(1-s^{2})(1-k^{2}s^{2})}}
  =
  \frac{1}{\sqrt{a}}
  \mathsf{K}(k),
\end{equation}
\begin{equation}
  I_{1}
  =
  -\int_{0}^{1}
  \frac{ds}{\sqrt{1-s^{2}}}
  =
  -\frac{\pi}{2},
\end{equation}
\begin{equation}
  I_{2}
  =
  \sqrt{a}
  \int_{0}^{1} 
  ds 
  \sqrt{
    \frac{1-k^{2}s^{2}}{1-s^{2}}
  }
  =
  \sqrt{a}
  \mathsf{E}(k),
\end{equation}
\begin{equation}
\begin{split}
  I_{3}
  &=
  -a 
  \int_{0}^{1}
  ds 
  \frac{1-k^{2}s^{2}}{\sqrt{1-s^{2}}}
  =
  -\frac{\pi a}{2}
  \left(
    1
    -
    \frac{k^{2}}{2}
  \right),
\end{split}
\end{equation}
\begin{equation}
\begin{split}
  I_{4}
  &=
  a^{3/2}
  \int_{0}^{1}
  ds 
  \sqrt{
    \frac{(1-k^{2}s^{2})^{3}}{1-s^{2}}
  }
\\
  &=
  -\frac{a^{3/2}}{3}
  \left[
    (1-k^{2})
    \mathsf{K}(k)
    -
    2 (2-k^{2})
    \mathsf{E}(k)
  \right],
\end{split}
\end{equation}
where $\mathsf{K}(k)$ and $\mathsf{E}(k)$ are the first and second kinds of complete elliptic integral. 
Substituting these formulas into Eqs. (\ref{eq:integral_s}) and (\ref{eq:integral_alpha}), 
Eq. (\ref{eq:ju_in_plane_def}) becomes 
\begin{equation}
\begin{split}
  &
  j(E)
  =
  -\frac{16 \alpha eM^{2}d}{ 3\hbar \eta} \times 
\\
  & 
  \frac{[-a^{2}(1-k^{2})+12(\epsilon^{2}-h^{2})] \mathsf{K}(k) + 2a[a(2-k^{2})-6(1-\epsilon)] \mathsf{E}(k)}
    {\sqrt{a} [4(1-\epsilon)-a(2-k^{2})]}.
  \label{eq:balance_current_u}
\end{split}
\end{equation}
In the limit of $E \to E_{\rm max}$ ($\epsilon \to (1+h^{2})/2$), 
Eq. (\ref{eq:balance_current_u}) gives $j_{\rm u-}$ in Eq. (\ref{eq:jc_max}). 
By changing the integral range, as mentioned above, $j_{\rm u+}$ is also obtained. 


Equation (\ref{eq:current_switch_lambda_in_plane}) is obtained in a similar manner. 
Equations (\ref{eq:integral_s}) and (\ref{eq:integral_alpha}) can be used to evaluate 
the integrals in Eq. (\ref{eq:current_switch_lambda}). 
Note that the integral range to derive Eq. (\ref{eq:jc_max}) is over an out-of-plane precession trajectory, 
while the range in Eq. (\ref{eq:current_switch_lambda}) is $[\mathbf{m}_{\rm d\pm},\mathbf{m}_{\rm d}]$. 
We notice that $a$ and $b$ in Eqs. (\ref{eq:integral_s}) and (\ref{eq:integral_alpha}) are 
$4h(1-h)$ and $b=0$ on the constant energy curve including the saddle point 
because $E_{\rm saddle}=MH_{\rm appl}$ ($\epsilon=h$). 
Then, Eqs. (\ref{eq:integral_s}) and (\ref{eq:integral_alpha}) for $j_{\rm th \pm}$ are given by 
\begin{equation}
  \mathscr{N}_{\rm s}^{\pm}
  =
  \mp
  \int 
  dm_{z}
  \frac{m_{z}^{2}-2(1-h)}{(1+\lambda m_{z}) \sqrt{4h(1-h) - m_{z}^{2}}},
\end{equation}
\begin{equation}
  \mathscr{N}_{\alpha}^{\pm}
  =
  \pm
  2\pi M 
  \int 
  dm_{z}
  \frac{[m_{z}^{2}-4(1-h)]m_{z}}{\sqrt{4h(1-h)-m_{z}^{2}}}.
\end{equation}
The integral range is $[2 \sqrt{h(1-h)},0]$ for $j_{\rm th+}$ and $[-2h\sqrt{h(1-h)},0]$ for $j_{\rm th-}$. 
We notice that $\mathscr{N}_{\rm s}^{\pm} = \mp [ J_{2}^{\prime} - 2(1-h) J_{0}^{\prime}]_{\pm 2 \sqrt{h(1-h)}}^{0}$ 
and $\mathscr{N}_{\alpha}^{\pm} = \pm 2\pi M [ J_{3} - 4(1-h) J_{1} ]_{\pm \sqrt{2 h(1-h)}}^{0}$, 
where 
\begin{equation}
  J_{n}
  =
  \int 
  dz 
  \frac{z^{n}}{\sqrt{a-z^{2}}},
\end{equation}
\begin{equation}
  J_{n}^{\prime}
  =
  \int 
  dz 
  \frac{z^{n}}{(1+\lambda z)\sqrt{a-z^{2}}}.
\end{equation}
Moreover, these integrals satisfy $J_{2}^{\prime} =(J_{0}^{\prime}-J_{0})/\lambda^{2} + (J_{1}/\lambda)$. 
Then, using the following formulas, Eq. (\ref{eq:current_switch_lambda_in_plane}) is obtained;
\begin{equation}
  J_{0}
  =
  \int 
  \frac{dz}{\sqrt{a-z^{2}}}
  =
  \sin^{-1}
  \left(
    \frac{z}{\sqrt{a}}
  \right),
\end{equation}
\begin{equation}
  J_{1}
  =
  \int 
  dz 
  \frac{z}{\sqrt{a-z^{2}}}
  =
  -\sqrt{a-z^{2}},
\end{equation}
\begin{equation}
  J_{3}
  =
  \int 
  dz 
  \frac{z^{3}}{\sqrt{a-z^{2}}}
  =
  -\frac{\sqrt{a-z^{2}}(2a+z^{2})}{3},
\end{equation}
\begin{equation}
  J_{0}^{\prime}
  =
  \int 
  \frac{dz}{(1+\lambda z) \sqrt{a-z^{2}}}
  =
  \frac{1}{\sqrt{1-\lambda^{2}a}}
  \sin^{-1}
  \left[
    \frac{z+\lambda a}{\sqrt{a}(1+\lambda z)}
  \right].
\end{equation}


\section{Instability condition in terms of magnetic field}
\label{sec:AppendixC}

In the main text, we derive the threshold current density as a function of the magnetic field. 
In some experiments [\onlinecite{suto12,bosu16,hiramatsu16}], on the other hand, the instability threshold is investigated 
by fixing the value of the applied current (voltage) and changing the magnetic field magnitude. 
The threshold magnetic field magnitude below which the self-oscillation is excited was found experimentally [\onlinecite{hiramatsu16}], 
which indicates that the threshold magnetic field is a decreasing function of $\theta_{H}$ ($0 < \theta_{H} \le 90^{\circ}$). 
The theoretical formula of the threshold magnetic field, $H_{\rm th}$, is, in principle, obtained 
by rewriting the instability threshold condition, Eq. (\ref{eq:current_switch}), in terms of the magnetic field. 
For example, when $\theta_{H}=90^{\circ}$ and $H_{\rm appl}/(4\pi M) \ll 1$, 
Eq. (\ref{eq:threshold_current_in_plane}) is rewritten as 
\begin{equation}
  H_{\rm th}(\theta_{H}=90^{\circ})
  \simeq
  4\pi M
  \left(
    \frac{\hbar \eta |j|}{16 e M^{2}d}
    -
    \alpha
    \sqrt{
      \frac{\hbar \eta |j|}{4 e M^{2}d}
      +
      4 \alpha^{2}
    }
    +
    2 \alpha^{2}
  \right).
  \label{eq:threshold_fild}
\end{equation}
Although it is difficult to derive analytical formula of the threshold magnetic field for an arbitary value of $\theta_{H}$ 
because the right hand side of Eq. (\ref{eq:current_switch}) is a complex function of the magnetic field, 
the experimental result [\onlinecite{hiramatsu16}] indicates that $H_{\rm th}(\theta_{H}) \sin \theta_{H} \simeq H_{\rm th}(\theta_{H}=90^{\circ})$. 


\section{Definition of the threshold current density in numerical simulation}
\label{sec:AppendixD}

We solve the LLG equation numerically from $t=0$ to $t=20$ ns 
by using the fourth-order Runge-Kutta method. 
The time step is $\Delta t=10$ fs. 
The threshold current density in the numerical simulation is defined as a minimum current density 
satisfying $|m_{x}(t=20{\rm ns})-m_{x}(t=20{\rm ns}-\Delta t)|>10^{-10}$ or $|m_{z}(t=20{\rm ns})|>0.9$, 
where the former means that the magnetization is in the oscillating state 
while the latter means that the magnetization moves to the $\pm \mathbf{e}_{z}$ direction. 





\end{document}